# Reallocation and Allocation of Virtual Machines in Cloud Computing

Manan D. Shah [*a], Harshad B. Prajapati [b]

[a] Department of Information Technology, Dharmsinh Desai University, Nadiad-387001- Gujarat, India.
[b] Department of Information Technology, Faculty of Technology, Dharmsinh Desai University, Nadiad-387001- Gujarat, India.
[*a] Email: - mananshah0003@gmail.com

*Abstract-*

*Cloud computing has given the new face to the distributed field. Two main issues are discussed in this paper, (I) "the process of finding the efficient virtual machine by using the concept of load balancing algorithm". (II) "Reallocation of the Virtual Machines" i.e. migration of the Virtual Machines when cloud provider is not available with the required Virtual Machines. We have discussed about the different load balancing algorithms which are used for deciding the efficient Virtual Machine for the allocation to the client on demand. While in the second issue is concern we have discuss about different modules available for the migration of Virtual Machines from one source machine to the other target machine. At last discussion about the different simulators available for the cloud are carried out in this paper.*

**1. Introduction**

Cloud is also known as the image of the internet. Thus while providing services accuracy is to be maintained. Cloud computing is a vast concept. The main interesting part carried out, is considered while migrating the VM and as per the second issue, numbers of the algorithms for the load balancing in the cloud computing are available for allocating the efficient VMs. Among such available algorithm which is to be used is the main decision is to be taken. Some of those algorithms have been discussed in this paper. Thus for having accurate usage of resources and being Faithfull with all the resources, concept of load balancing is being carried out. The main aim of cloud computing is to provide better service to the user as per their requirement.

1. To study the performance of the existing load balancing algorithms for allocating the efficient VMs on demand.

2. To design the modification of the existing load balancing algorithm for achieving the better usage of the services provided by the cloud.

We achieve such objectives by having the clear concept of the existing load balancing algorithms. Concept in the sense of better performance during providing resources, having better response time and many such useful concepts of the existing algorithms are discuss on the basis of the previous research done on existing load balancing algorithms.

Moreover we have discuss about cloud computing concept in the below section. After that we have discussed about research problem with its some available solutions, proposed algorithm, comparison of available simulators for the cloud computing and at last some implemented results.

**2. Background Theory**

One of the most on demand requirement of the 21st century computing is that users should have access to the internet over the portable devices rather than through some desktop pc. As the users won't have powerful machines. In such case cloud computing services are used from wide spread resources, rather than remote server or local machines. Let us make it clear the concept of cloud computing. Now suppose if I am an, End user I need an application s/w to do my job [A Computer]. As a Businessman I need an infrastructure to run my website such as server and if suppose I am a web application developer I need a perfect platform that suits me. In short (i) End user: - needs Application (ii) Businessman: - needs Infrastructure (iii) Web Developer: - needs Platform. While providing such services some drawbacks took place such as up gradations, server maintenance and many such things. Thus to overcome such drawback cloud computing took place. It is nothing but a Metaform of the internet.

Thus the standard definition given for the cloud computing by NIST is that, *"cloud computing is a model for enabling convenient, on demand network access to a shared pool of configurable computing resources (e.g., network, server, storage, applications and services) that can be rapidly provisioned and released with minimal management efforts or service provider interaction"* [1]. Thus the three main services provided by the cloud are Infrastructure as a service (IaaS), Platform as a service (PaaS) and Software as a service (SaaS).

Cloud computing architecture: - Cloud computing consist of four different layers, each layer having their own functionalities. Moreover the services provided by the cloud computing are also mentioned in the below figure. Let us have a look to all the four layers with the help of diagram.

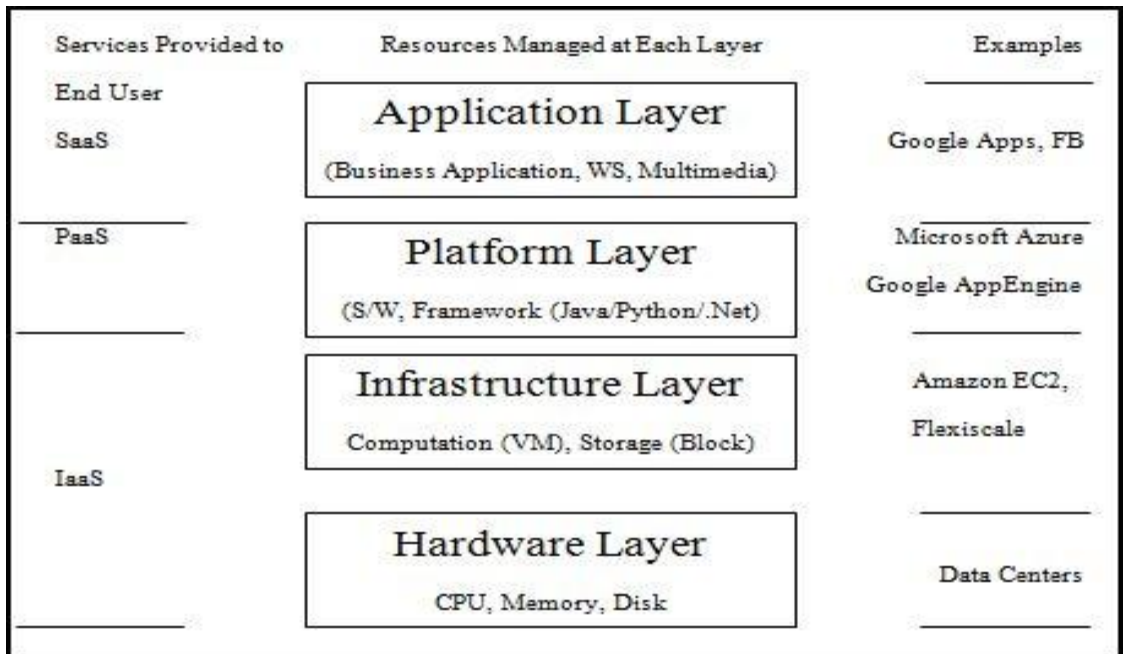

**Figure 1 Cloud Computing Architecture**

Above mentioned figure is showing the four different layers of the cloud computing, they are hardware layer, infrastructure layer, platform layer and application layer.

Cloud computing is also known as service-driven business model. In other words, hardware and platform level resources are provided as services on an on-demand basis. Basically, every layer of the architecture is implemented as a service to the layer above. However, clouds provide services that are grouped into three categories: software as a service (SaaS), platform as a service (PaaS), and infrastructure as a service (IaaS).

Software as a service: - It plays a role while providing services on demand over the internet. The main common live examples of SaaS are Google App and facebook. Different applications are provided by the different servers through the internet which are used as a service to the users. While using the software user has no need to make lots of changes or doesn't require integration to other system.

Platform as a service: - PaaS provides all the resources that are required for building applications and services completely from the Internet, without downloading or installing software [1]. The common examples of PaaS are Google AppEngine and Microsoft Azure. Different services provided by the PaaS are software design, development, testing, deployment, and hosting. Other services are team collaboration, database integration, web service integration, data security, storage and versioning etc. PaaS refers to providing platform layer resources, including operating system support and software development frameworks.

Infrastructure as a service: - It is also known as Hardware as a service (HaaS). The live common example of IaaS is Amazon EC2. It offers hardware as service to an organization so that it put anything into the hardware as per the requirements of the users. HaaS allows the user to purchase resources on rent as server space, network equipment, memory, CPU cycles and storage space. User will have a virtual machine as per their required configuration by providing their required input to the cloud provider.

2.1 Types of Cloud

As per the different requirements of the different users, cloud is being categorized into 3 different types of clouds they are, Private cloud, Public cloud and Hybrid cloud.

Private Cloud: - A cloud for the private organization can be considered as a private cloud. Now suppose if a college having their own cloud then such a cloud is considered as a private cloud. It is also known as internal clouds; a private cloud may be built and managed by the organization or by external providers. A private cloud offers the highest degree of control over performance, reliability and security. An example: - A cloud of a single organization.

Public Cloud: - A cloud for more than one organization is known as a public cloud. Now suppose if the cloud has being created for the university which includes numbers of colleges then such a cloud is considered as public cloud. A cloud in which, service providers offer their resources as services to the general public. Service provider gets several benefits such as no initial capital investment on infrastructure and shifting of risks to infrastructure providers. For E.g. A cloud for more than one organization.

Hybrid Cloud: - Combination of both the private as well as public cloud is known as hybrid cloud. In a hybrid cloud, part of the service infrastructure runs in private clouds while the remaining part runs in public clouds. Hybrid clouds offer more flexibility than both public and private clouds.

**3. Research Problem**

As the main aim of cloud computing is to provide resources as a service on demand to the user. In this research paper we are going to carried out two main problems.

1. Reallocation of virtual machines.
2. Allocation of virtual machines.

*3.1 Reallocation of virtual machines*

As we know the main aim of cloud computing is to fulfill the requirement of the user by providing resources as a service. Now suppose if we are having two physical machines and both the machines having one virtual machine with the processor of 1 GHz. If the request arrives for the virtual machine having processor of 2 GHz, at that time the concept of reallocation of virtual machine took place. For such situation we migrate one of the virtual machine from one physical machine to the other and combining configuration of both the virtual machine and make it available with the configuration of 2GHz and should provide a better service to the user.

It is possible to have a number of cloud provider, if the situation took place that the request arrive at one of the cloud provider and somehow that cloud is not available with the resource required by the user, so in such case the other cloud provider, migrate their resources to that cloud and fulfill the requirement of the user.

*3.2 Allocation of virtual machines*

Now while allocating the virtual machines to the user as per the requirement, it is also need to be kept in mind that we should allocate them an efficient virtual machine. When the user request arrives for the virtual machine and if the cloud provider is available with the number of free virtual machines, then allocating an efficient virtual machine is the main goal of cloud provider.

This problem contains the issue of allocating efficient virtual machines by using the concept of load balancing algorithm. For e.g. if the request arrives at the cloud provider for the VM and if more than 1 virtual machines are available, then the load balancing algorithm will find an efficient VM for the allocation. Moreover if arrived request does not match with the configuration of the available virtual machine then we disclose the suitable virtual machines configuration to the user so that they able to decide whether they need to remain in queue or get service from other cloud service provider.

**4. Existing Solutions**

*4.1 Existing Load Balancing Algorithms*

In complex and large system there is a tremendous need for load balancing. For simplifying load balancing globally (e.g. in cloud), employing techniques would act at the components of the cloud in such a way that the load of the whole cloud is balanced. As in our case while during the allocation of the VMs the load balancing algorithms plays an important role. As when the request arrives at the data center, and to decide which VM is to be allocated, will be decided with the help of the load balancing algorithms. Thus different load balancing algorithms for the cloud computing are discussed here.

1. Round Robin Algorithm
Round robin algorithm is random sampling based. It means it selects the load randomly in case that some server is heavily loaded or some are lightly loaded [5]. It randomly selects virtual machines for the allocation. This algorithm affects, if the more than one virtual machine is freely available and if, on randomly bases the virtual machines are going to be selected, then possibility of selecting more configured virtual machine as compared to the demanded machine will be high.

## 2. Equally Spread Current Execution Algorithm

Equally spread current execution algorithm process handle with priorities. It decides the efficient virtual machine on the bases of priority. It distribute the load randomly by checking the size and transfer the load to that virtual machine which is lightly loaded or handle that task easy and take less time , and give maximize throughput [5]. It is spread spectrum technique in which the load has been divided between multiple virtual machines.

## 3. Active Monitoring Load Balancer

Active VM Load Balancer maintains information about each VMs and the number of requests currently allocated to which VM. It will check the available virtual machines. If there are more than one, the first identified is selected. Data center controller will receive an id of the virtual machine and it will send the request to the VM identified by that id. Active VM Load Balancer will be notifies of the new allocation.

## 4. Throttled Load Balancing Algorithm

Throttled algorithm is completely based on virtual machine. In this client first requesting the load balancer to check the right virtual machine which access that load easily and perform the operations which is give by the client or user[8]. In this algorithm the client first requests the load balancer to find a suitable Virtual Machine to perform the required operation. Moreover it has been proved by many researchers that from the number of available load balancing algorithm, it is proved that throttled load balancing algorithm is better as compare to the other existing load balancing algorithm.

### *4.2  Migration Modules*

Now when should we take decision for migrating virtual machines? If the cloud provider receives request for the virtual machine and somehow if the cloud is not able to provide the service to user, then the other cloud provides VM to the cloud which is providing service to user. Some modules which are needed to be taken in consideration while migrating the virtual machines are discussed below.

1. Migration Decision Maker

2. Migration Controller

3. Resource Reservation Controller

4. Resource Monitor

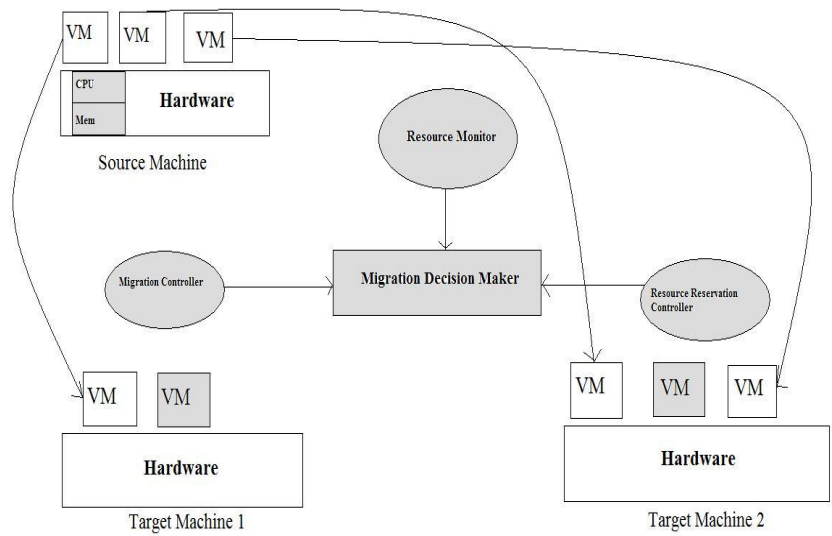

**Figure 2: - Resource Reservation Technology for Live Migration Framework of Multiple Virtual Machines. The VM in shaded form represents the reserved resources in the form of virtual machines in the target machines, the CPU and Memory in shaded represents the reserved resources in the source machine [Adapted from [15]].**

1. Migration Decision Maker

It plays most important role in the live migration framework of multiple virtual machines. For making an effective migration decision it plays a most important role. Quality of this module is measure by the effectiveness and efficiency of migration strategies. While using a machine learning methods this module also have some intelligence to make efficient migration strategies. Currently, some simple migration strategies such as sequential migration, parallel migration and workload aware are included in this module.

2. Migration Controller
The total management of migration from source machine to the target machine is controlled by this module. It controls the real migration process. It will choose the right target machine from the candidate target machine list and trigger the migration at particular time. This module relies on the Migration Decision-Maker module, and executes the migration strategies made by the Migration Decision-Maker module. It takes care for not occurring failure during the migration.

3. Resource Reservation Controller
It implements different resources reservation strategies for both source machine and target machine, such as CPU resource and memory resource reservation in the source machine or the whole virtual machine resource reservation in the target machine. This module is very important in the live migration of multiple virtual machines to avoid migration failures because of the insufficient resources in the target machine for the migrated virtual machine. The resource reservation in the source machine is implemented by dynamically adjusting the migrated virtual machine's CPU cycle and the memory resource. And the resource reservation in the target machine is implemented by temporarily creating particular virtual machines to occupy a certain number of system resources for the migrated virtual machines [15].

4. Resource Monitor
It maintains the records of the resources which are reserved and which are free. It is responsible for monitoring the resource status of both virtual machines and physical machines, including the resource utilization, virtual machine configuration information which is essential to make migration decisions. It is also used to analyze the workload stability which is useful to avoid migration thrashing.

## 5. Proposed work

*5.1 Proposed Algorithm: -*

While doing some modification to the existing load balancing algorithm we achieve our goal. While displaying the available virtual machines with their configuration to the user, so that the user able to decide from which service provider they should take a service.

Below discussed are the steps followed by doing modification to the available Throttled load balancing algorithm.
Steps followed by the proposed algorithm.

1. Efficient algorithms find expected efficient Virtual Machine for the allocation.

2. When a request to allocate a new VM from the Data Center Controller arrives, Algorithms find the most efficient VM for allocation from the bunch of available VMs.

3. Efficient algorithms return the id of the efficient VM to the Datacenter Controller.

4. Available virtual machine id will be disclosed to the end user.

5. If the required VM is not available, efficient algorithm will suggest some related VMs to the user.

6. When the VM finishes processing the request and the Data Center Controller receives the Response. Data center controller notifies the efficient algorithm for the VM de-allocation .Continue From Step 2.

Once the request arrives with the configuration of the virtual machine, algorithm will find out the matching configuration of the virtual machine with the available virtual machines. It will display the VM ID to the user if the configuration matches with the available VM and if the required VM is not available then the algorithm will disclose some relevant VMs with their configuration. For better understanding results are shown in below section.

## 6. Methodology

*6.1 Simulators: -*
The main aim of simulator is to test the implementation work in the absence of the required environment. Thus in the cloud environment two simulator are used CloudSim and Vcloud. CloudSim is the open source. Some simulators available for the distributed field such as SimGrid, GridSim, etc such simulators are not valid for the cloud computing as the cloud environment having multiple layers while SimGrid and GridSim are made for the single layer environment. Some comparison of simulator available for cloud computing are discussed below.

TABLE I. COMPARISON OF SIMULATORS

| | **CloudSim** | **Virtual Cloud** |
|---|---|---|
| Physical Machine | The allocation policy is applied at initial state. In case of storage, cloudsim has partially implemented storage area network. All the files needed to start an application is first copied & the application starts. | While in Virtual Cloud operations are performed simultaneously while the application is running. If case of storage area networks is considered then all the disk related operations will be performed during the application is running. |
| Communication Between VMs | In the cloudsim, all the communication between virtual machines happen, before the start of the application. | While in Vcloud communication between VMs are designed, to send/receive packets. While during the communication between, two VMs on different physical machines, the send and receive packets are control by the hypervisor. So load on the hypervisor increases. |
| Migration Policies | Currently, cloudsim implemented a migration policy based on load on a specific physical machine. | While in Vcloud manager module is to be design, which will implement these policies, based on the current resource utilization at the virtual and physical machine. |
| Scalability | Currently cloudsim uses configuration about physical and virtual machines, using java source code. | While Virtual Cloud uses XML file for representing the configurations of physical and virtual machines. |
| Resource | In cloudsim approach, there is no interference between different resources. Here interference means, usage of disk operation, also creates CPU load. | While in Virtual Cloud, we have considered interference among the different resources. |

All the available load balancing algorithms are being tested in the cloudsim simulator. Moreover cloudsim is an open source. Thus we are going to use CloudSim for our experimental testing.

**7. Experimental results**

*7.1 Results of Allocation of VMs*

Users are going to decide VMs configurations as per their requirements. In our case user are going to decide VMs configuration on the bases of RAM, No. of CPUs and the Space required to the user. Below figure shows the basic image of input taken from the user.

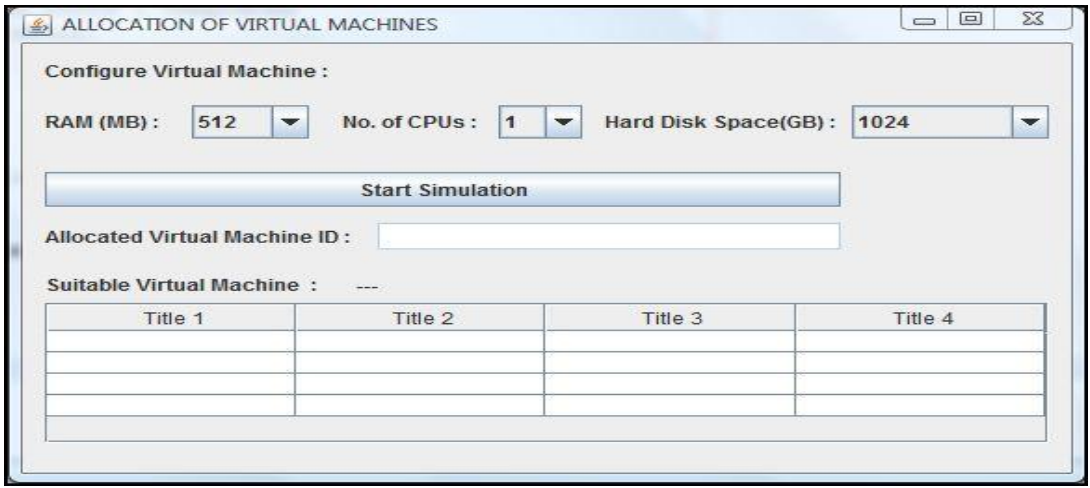

**Figure 4: - Input of required RAM, No. of CPUs and Hard Space are given by user.**

Case 1: - Let us say if the user enters configuration of VM with the requirement of 512MB RAM, No. of CPUs = 1 and Space required = 1024GB. Output is as shown in below figure.

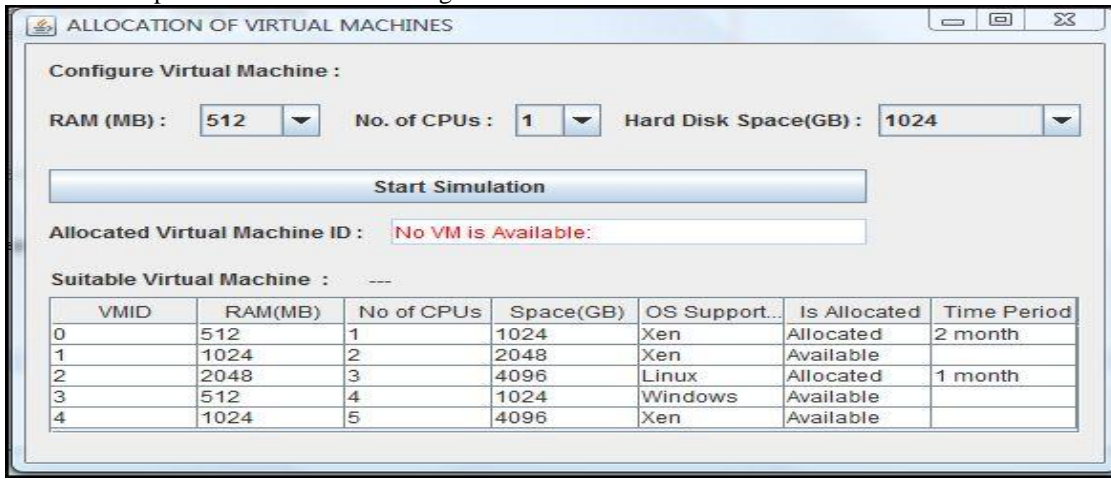

**Figure 5: - Result 1(required VM is not available)**

Above mentioned output shows that the virtual machines with the configuration of 512 MB RAM, 1 CPU and 1024 GB Hard Disk Space is already allocated to the other user for 2 months so it will display that no VM is available as per the required configuration, but it shows some of the VMs to the user to decide as per their requirement.

Case 2: -Let us say if the user enters configuration of VM with the requirement of 812MB RAM, No. of CPUs = 1 and Space required = 1024GB.
Now suppose if we don't have the VM available with the required configuration of the user, so in that case we should display the suitable VM configuration available with us to the user. Output will be as shown in below figure.

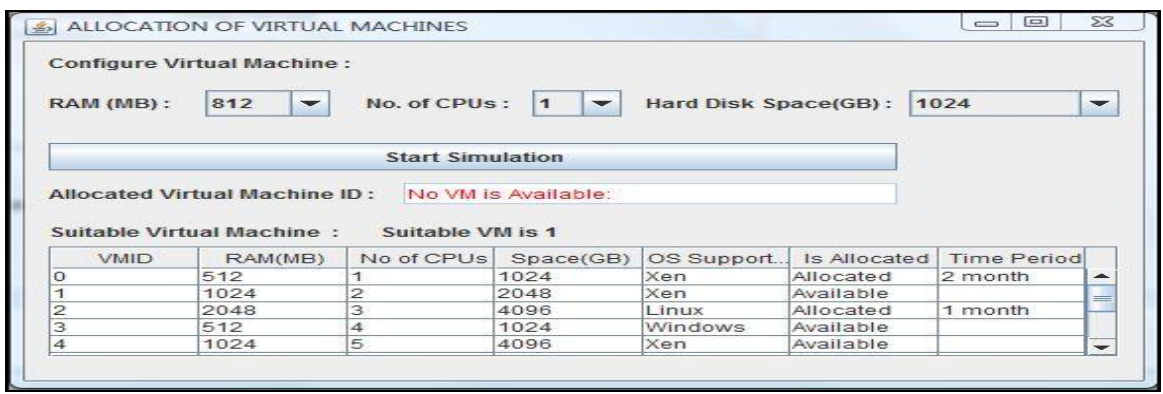

**Figure 6:- Result 2(Suggest Some VM if required configured VM is not available)**

Case 3: - Let us say if the user enters configuration of VM with the requirement of 1024MB RAM, No. of CPUs = 5 and Space required = 4096GB. Output is as shown below.

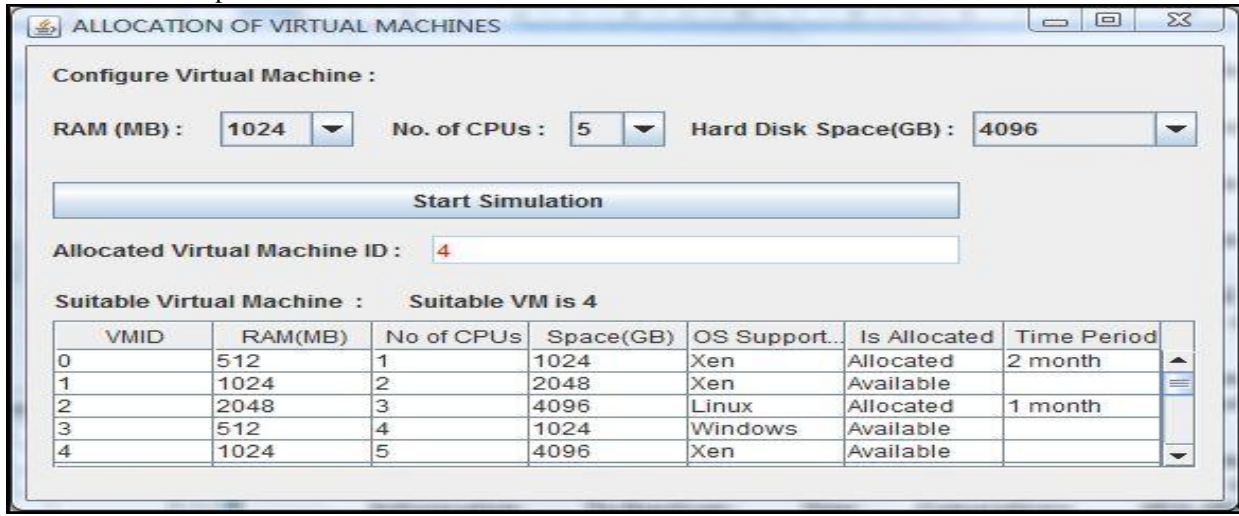

**Figure 7:** - Result 4(Display Allocated VM ID)

**Conclusion**

We proposed an algorithm for finding out the efficient VM for the allocation as per the user's requirement. In this paper, we discussed about the some existing load balancing algorithms which are used for allocation of efficient virtual machine. Our proposed algorithm is used for handling request, which arrives for the requirement of VMs. We suggest some of VMs with their configuration, if the required VM is not available. Cloudsim is a simulator, which we used to test our algorithm. In future we wish to add new steps in our proposed algorithm which will able to provide better solution for getting better service from the cloud provider.